# Micro-Tensile Tests on Micromachined Metal on Polymer Specimens: Elasticity, Plasticity and Rupture

C. Seguineau[1-2], M. Ignat[2], C. Malhaire[3], S. Brida[4], X. Lafontan[1], J.-M. Desmarres[5], C. Josserond[2], L. Debove[2]

1 Nova MEMS, 10 av. de l'Europe, F-31520 Ramonville, France
(Ph : +33(0)5 6128 1935, cedric.seguineau@novamems.cnes.fr)
2 ENSEEG INPG, SIMAP UMR 5266, Université Joseph Fourier, Saint-Martin-d'Hères, F-38402, France
3 Université de Lyon, INSA-Lyon, INL, CNRS UMR 5270, Villeurbanne, F-69621, France
4 AUXITROL S.A. , Esterline Sensors Group, Bourges, F-18941, France
5 CNES, DCT/AQ/LE, bpi 1414, 18 Avenue Edouard Belin, 31401 Toulouse Cedex 9, France

*Abstract-* **This study is focused on the mechanical characterization of materials used in microelectronic and micro- electromechanical systems (MEMS) devices. In order to determine their mechanical parameters, a new deformation bench test with suitable micromachined specimens have been developed. Uniaxial tensile tests were performed on "low cost" specimens, consisting in electroplated thin copper films and structures, deposited on a polimide type substrate. Moreover, a cyclic mechanical actuation via piezoelectric actuators was tested on the same deformation bench. These experiments validate the device for performing dynamic characterization of materials, and reliability studies of different microstructures.**

## I. INTRODUCTION

Submicron thin films are essential elements of microelectronic and MEMS devices. However, it is already well established that their mechanical properties generally differ from their bulk material counterpart, and that these properties depend strongly on their microstructure. Consequently, there is a permanent lack of information on the mentioned parameters, insofar as microstructure depends on the fabrication conditions which vary from a supplier to another. Insofar as their mechanical behaviors may affect the performances and the reliability of the devices, improving the knowledge of the mechanical properties of thin films remains one of the main issues encountered in the development of MEMS. Particularly, the device lifetime may be shorten after repeated thermal, electrical or mechanical cycles. By analysing their mechanical response, then identifying the degradation mechanisms, solutions may be found directly related to design and/or microstructural changes in the materials.

The main purposes of our work are : first to determine the mechanical and electrical properties of thin film materials, second, to discuss the results with respect to microstructural analysis (XRD, EBSC, TEM,...) and third, to analyse the correlations between physical and micromechanical determined parameters. In this contribution, we present some examples of the mechanical testing of thin films on substrate systems which involve electroplated copper on a polimide type substrate. From our micro-tensile experiments, elasticity, plasticity and rupture are discussed.

The experimental approach is based on a new micro-tensile bench test and the development of suitable micromachined specimens. We may note here, that to avoid any substrate effect [1-3],we also performed some previous experiments on self-standing thin films [4].

## II. EXPERIMENTAL DESCRIPTION

### A. Micro-tensile bench test

The micro-tensile testing machine, shown in Fig. 1, consists in a gripping device actuated by a DC micro-motor. A laser displacement sensor measures the relative displacement of the two movable grips. Forces are measured using a miniature piezoresistive load cell (RDP Electronics Ltd.). Moreover, two piezoelectric transducers make deformation cycles for fatigue testing. This experimental setup allows cycles up to a few hundred of Hertz.

The schematic view of a typical test specimen and the tensile principle are shown in Fig. 2. The thin films have a beam shape, located in a silicon frame insuring the specimen integrity when mounting the sample. Once the latter fixed on the grips of the test bench, the frame is cut, releasing the free-standing beam. During the test, the DC micromotor drives the "jack" and the two grips are separated at a constant displacement rate, pulling on the sample.

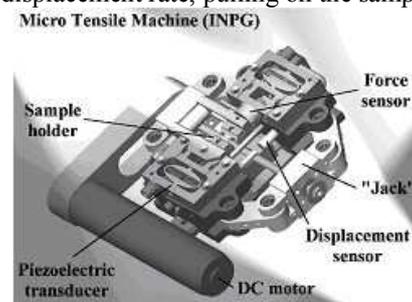

Fig. 1. Schematic view of the micro tensile machine

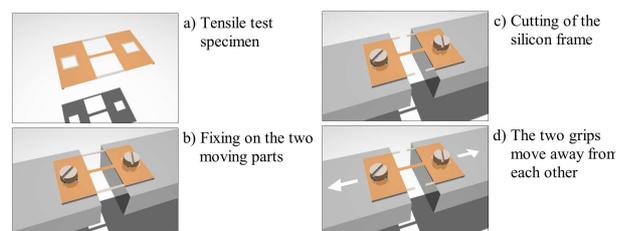

Fig. 2. Specimen fixation procedure (artist view)

*B. Samples description*

Two different technologies have been chosen to fabricate the specimens: silicon standard micromachining process and metal on polymer process. We have first developed silicon frames sustaining submicron thin self-standing films. The test structure design have been optimized using Finite Element Simulations. The feasibility of SiN and Al micro-tensile specimen with very large length over thickness ratio using a standard process has been demonstrated [4]. However, because approaching the downscaling limits of standard tensile testing, thicker samples of a copper electrodeposited on Polimide were shaped for tensile tests by laser cutting. A total of 128 specimens have been achieved on a 50µm thick A4 polymer sheet. Some uncut samples are shown in Fig. 3.

The 8.5µm thick copper film was previously obtained by electrodeposition and patterned. Four kinds of samples have been achieved and studied: bare Polimide beams, fully coated samples by the copper film, and beams with patterns of copper lines or round and rectangular structures, as shown in Fig. 4. These metal patterns can be used for *in-situ* resistivity measurements and be observed by microscopy during and after the tensile test to determine local strains and plastic deformations. Some uncut samples are shown in Fig. 3.

The tensile sample's dimensions were: 5mm overall length with 3mm gage length and 500µm width. The front side metal layer was patterned by a standard photolithography and etching process before laser cutting. Only the polymer film was cut around the metal patterns so we can assume that the metal film did not suffer from heating.

### III. RESULTS AND DISCUSSION

*A. Micro-tensile experiment*

A result for a Cu-polymer specimen is given in Fig. 5. In the initial portion of the force-displacement curve, a linear behaviour is observed. The overall response is typical of a very ductile material up to fracture. After a six load/unload cycle only, a fatigue damage process appears: after each load/unload cycle, the apparent elastic modulus decreases as shown in Fig. 6. The analysis of such experimental results with the law of mixture formalism is currently developed. As a matter of fact, awkward deformations of the micro-tensile device alter the first loading steps (elastic response of the specimen), leading to a misevaluation of Young's modulus. The bench response at shallow load is currently analysed and will be taken into account in the Young's modulus evaluation. However, we are confident about stress level. Thus, these samples show tensile flow and rupture strengths of about: 160MPa and 350MPa, and a total elongation of about 2µm.

*B. Cycling with piezoelectric actuators*

Moreover, a second kind of experiment has been made, with a cyclic piezoelectric actuation. Increments of tensile load have been applied by the DC micromotor, with holding plateau between them. On each plateau, cyclic solicitations ($f$=100Hz, d=30µm) have been imposed by the piezoelectric transducers. Several points may be noted. First, the load cell average value tends to zero rather quickly, proving that the stress level in the polymer relaxes during the cyclic solicitations. Second, the amplitude of the resulting displacement also tends to zero. These two observations are linked with the viscoelastic behaviour of the polymer. At last, DC micromotor was actuated with the piezoelectric cyclic solicitations, fixed at 350Hz. Then, the load rate of the DC motor was slight enough to ensure stress relaxation induced by the piezoactuators. Thus, the onset of local damage on the sample was observed in-situ with an optical microscope. As reported on Fig. 7., cracking is nucleating at the circular copper/polimide interfaces singularities, where the tensile stress components are the highest. Then the cracks propagate aroutn the circular pattern, by following the cyclic loading. Final failure of the samples is in general (3 experiments until now) associated to the bigger circular pattern.

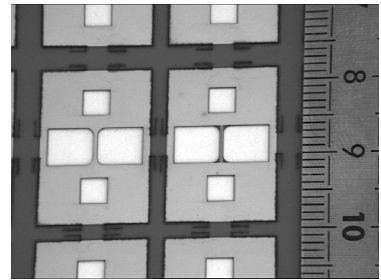

Fig. 3. Cu on polymer specimens

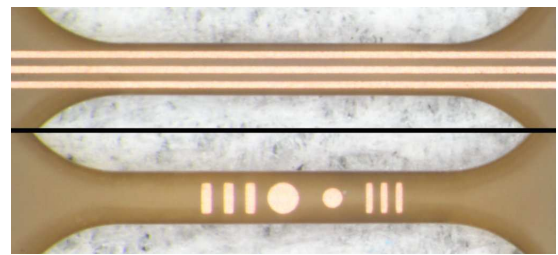

Fig. 4. Different geometries of copper patterns on the polimide substrate, before straining.

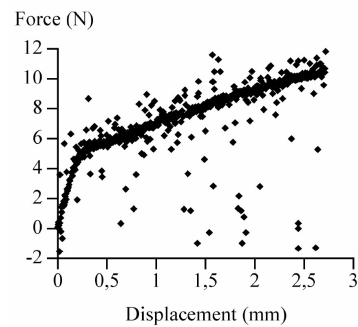

Fig. 5. Force-Displacement curve for a Cu/polymer specimen.

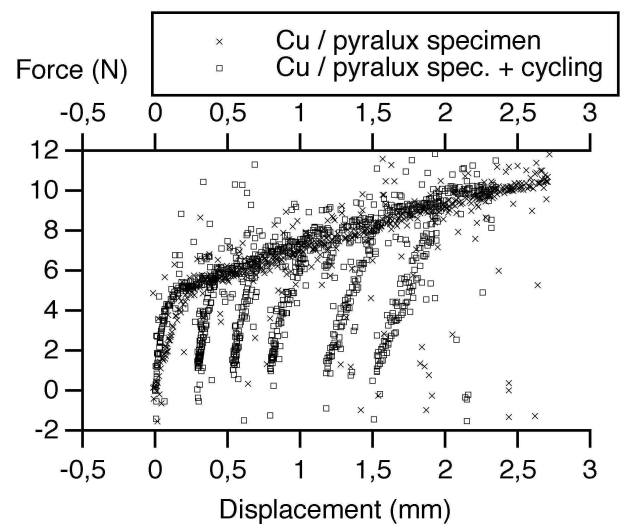

Fig. 6. Force-Displacement curve for a Cu/polymer specimen and cycling.

## IV. CONCLUSION

Micro-tensile experiments are useful to determine the mechanical response of thin films and coatings used in MEMS and most microelectronic devices. We have determined some mechanical properties of copper on polymer samples: specifically, the tensile flow and the rupture strengths. They are in agreement with the expected order of magnitude just like the large strains we obtained, usual for polymer materials. Eventhough our micro-tensile tester already allows to determine apparent values of parameters related to the elastic response of the tested materials (Young's modulus), further analysis of the tester compliance is needed to avoid awkward effects. Besides, an evolution of the apparent modulus with the loading cycles of a copper on polymer sample has been detected, certainly linked with the damage of the specimen, which can be induced because of the strain incompatibilities among the materials constituting the sample, at high induced strains. As expected, for samples with different geometries of patterned copper, the cracks nucleate at singular zones of stress concentrations, located at the copper/polymer interfaces. The piezoelectric additional stage has been successfully tested too and will be useful for future reliability studies

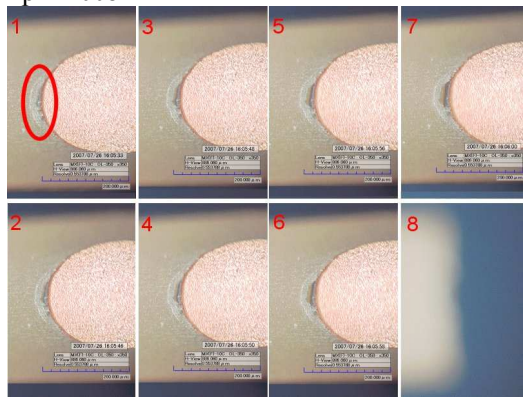

Fig. 7. Failure of a Cu/polymer specimen under piezoelectric cyclic actuations (350Hz)